\documentclass{elsart}
\usepackage{natbib}
\usepackage{epsfig}
\begin{document}
\runauthor{Colin}
\begin{frontmatter}
\title{Measurement of air-fluorescence-light yield induced by an electromagnetic shower }

\author[LAPP]{P. Colin}\author[UTAH]{}\footnote{contact address: {\tt colin@mppmu.mpg.de}},
\author[JINR]{A. Chukanov}, \author[JINR]{V. Grebenyuk},
\author[JINR]{D. Naumov},
\author[IPNL]{P. N\'ed\'elec}, \author[JINR]{Yu. Nefedov},
\author[LIP]{A. Onofre}, \author[JINR]{S. Porokhovoi}, \author[JINR]{B. Sabirov,}
\author[JINR]{L. Tkatchev}

\textbf{The MACFLY Collaboration:}

\address[LAPP]{Laboratoire d'Annecy-le-Vieux de Physique des Particules, IN2P3/CNRS, \\  Universit\'e de Savoie, Annecy-le-Vieux, FRANCE}
\address[IPNL]{Universit\'e de Lyon, Universit\'e  Lyon 1, CNRS/IN2P3/IPNL\\ Lyon, FRANCE}\address[JINR]{Joint Institute for Nuclear Research \\ Dubna, Moscow region, RUSSIA}
\address[LIP]{Laboratorio de Instrumenta\c{c}\~{a}o e Fisica Experimental de Particulas\\ Coimbra, PORTUGAL}
\address[UTAH]{Now at the Max-Planck-Institut f\"{u}r Physik, Munich, GERMANY}

\begin{abstract}
For most of the Ultra-High-Energy-Cosmic-Ray (UHECR) experiments
and projects (HiRes, AUGER, TA, JEM-EUSO, TUS,...), the detection
technique of Extensive Air Showers is based, at least, on
the measurement of the air-fluorescence-induced signal. The
knowledge of the Fluorescence-Light Yield (FLY) is of paramount
importance for the UHECR energy reconstruction. The MACFLY
experiment was designed to perform absolute measurements of the
air FLY and to study its properties. Here,
we report the result of measurement of dry-air FLY induced by 50\,GeV
electromagnetic showers as a function of the shower age and as a
function of the pressure. The experiment was performed at CERN
using a SPS-electron-test-beam line. The result shows the air FLY is
proportional to the energy deposited in air ($E_d$). The ratio
$FLY/E_{d}$ and its pressure dependence remain constant
independently of shower age, and more generally, independently of
the excitation source used (single-electron track or air shower).
\end{abstract}
\end{frontmatter}

\section{Introduction}
The most challenging research in the field of cosmic-ray physics
is certainly the highest-energy region of the cosmic-ray spectrum
(around $10^{20}$\,eV) where the GZK effect~\cite{GZK} is predicted.
The origin of these Ultra-High-Energy Cosmic Rays (UHECR) is still
enigmatic. Precise measurements of their spectrum and arrival-direction
anisotropy could provide important clue to understand their origin.
However, the UHECR are rare, and their detection is an experimental challenge.
In the 60's, a measurement technique based on the air-fluorescence light
was proposed. This light, observed in the
near-UV region ($\sim$300-400 nm), is induced by the de-excitation
of the air molecules (mainly N$_{2}$) occurring along the
Extensive Air Showers (EAS) produced by the interaction of UHECR with
the atmosphere.
Since, most of the past, current and future experiments:
HiRes~\cite{HiRes}, Pierre Auger Observatory~\cite{Auger}, Telescope
Array~\cite{TA}, Ashra~\cite{ashra}, OWL~\cite{owl}, EUSO/JEM-EUSO~\cite{euso},
TUS~\cite{tus}, use the air-fluorescence signal to
measure the UHECR flux, the knowledge of the Fluorescence-Light Yield (FLY)
is of paramount importance for the UHECR energy reconstruction.

In the past, the air fluorescence was a subject of extensive laboratory measurements.
In 1967, A. N. Bunner summarized the existing data in his thesis~\cite{Bunner}
and proposed a FLY model with an uncertainty estimated at $\sim$30~\%.
In spite of electron-beam-based measurements of Kakimoto et al. in 1996
\cite{Kakimoto} and Nagano et al. in 2003 \cite{Nagano}, the
uncertainties were still large inducing important systematics to
UHECR experiments. A historical review of the air-FLY measurements
can be found elsewhere \cite{history}.

The controversy and discrepancy between the AGASA~\cite{agasa} and
HiRes~\cite{HiRes} experiments lead the community to pursue its
effort to improve the knowledge of the air fluorescence (absolute light yield, spectrum)
and of its dependencies with pressure, temperature, humidity, electron energy,
shower age, etc. Since 2002, a dozen of new independent experiments were
proposed and took data. In these experiments, the air is excited with
a high-energy-electron beam (like for MACFLY, FLASH~\cite{flash1}~\cite{flash2} and AirFLY~\cite{airfly}), a radioactive source \cite{airlight} \cite{paris} \cite{alpha}
or a low-energy-electron gun~\cite{egun}. A comparison and discussion of the results
of these experiments can be found in the proceedings of the latest air-fluorescence workshop~\cite{summary}.
Two types of experiments were carried out, using thin or thick target.
For the first type the fluorescence light is induced directly
by interactions of primary electrons with the air,
whereas for the second type, the light is induced
by interactions of electromagnetic showers similar to EAS.
Only two experiments used thick targets: FLASH~\cite{flash2} and MACFLY.
%\footnote{An overview of the FLY experiments
%presented at the third workshop on air fluorescence -
%IWFM05~\cite{IWFM} should be found in~\cite{nedelec}.}

The MACFLY (Measurement of Air Cherenkov and Fluorescence Light
Yield) experiment has been designed to measure the light
induced by both a single-electron track and a high-energy
electromagnetic shower developing in air~\cite{Colin}.
Actually, the experiment is composed of two devices, MF1 for the
single-track fluorescence (thin target) and MF2 for the
electromagnetic-shower fluorescence (thick target). The MF1 results,
obtained for different electron energies (1.5\,MeV, 20\,GeV and 50\,GeV),
as well as an air-FLY model were previously published~\cite{MF1}.
In this paper, we focus on the air-FLY measurements performed
with the MF2 device at the CERN-SPS-X5-electron-test-beam line with
50\,GeV electromagnetic showers. We compare these measurements
with Monte Carlo simulations of shower development implementing a FLY model
based on the MF1 results.

\section{Experimental setup}
\begin{figure}
  \includegraphics[width=\columnwidth]{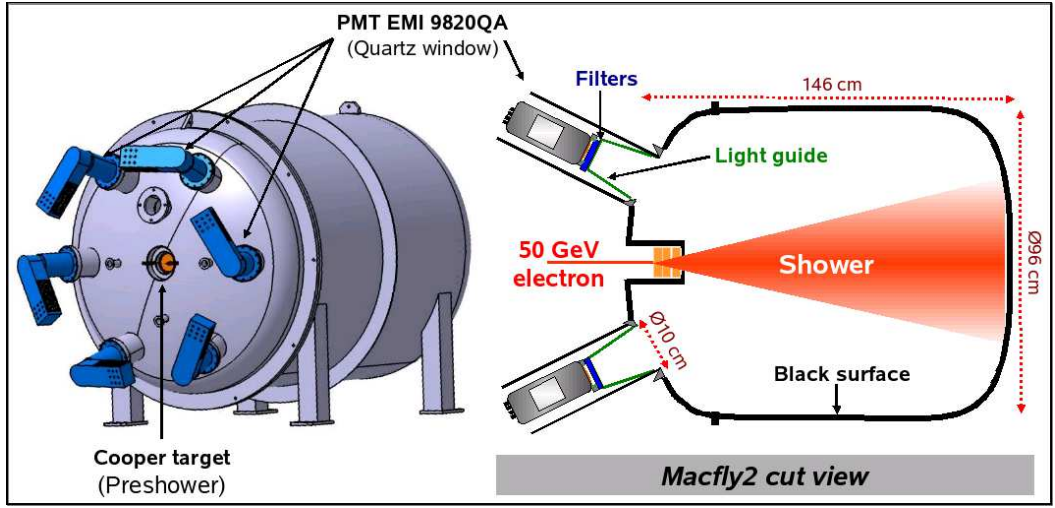}
  \caption{The MF2 chamber: schematic view (left) and cut view (right).}
  \label{fig-MF2_overview}
\end{figure}

The MF2 device is composed of a pressurized chamber containing the gas under study,
a pre-shower system and an optical system.
The chamber is a quasi-cylindrical (96\,cm in diameter and 146\,cm long) large-volume ($\sim$\,1\,m$^3$) tank internally covered with black paper
(see Figure~\ref{fig-MF2_overview}).
The electron beam is aligned with the axial symmetry axis of the tank.
It impinges on the pre-shower system which is a variable thickness target
used to initiate electromagnetic showers inside the chamber.
The pre-shower system is installed upstream the chamber, in a recess
of 15\,cm deep from the entrance wall of the tank (end-cap).
The optical system measures the fluorescence light emitted by the
excited air contained in the tank. It uses six UV-sensitive phototubes (PMTs)
EMI9820QA. These PMTs are installed on the entrance end-cap on a 350\,mm
radius circle centered on the beam line. They point to the gas volume making
a $20.5^{o}$ angle with the beam direction.
They are separated from the inner gas by quartz windows.
The optical system is also composed of Winston cones and of different colored
glass filters. The results presented in this paper correspond to
the measurements performed using the Schott BG3 filters, which has
a large transmittance band (290-440\,nm). The same filter was used
for the MF1 measurements (The MF1 device is described elsewhere~\cite{MF1}).

\begin{figure}
  \includegraphics[width=\columnwidth]{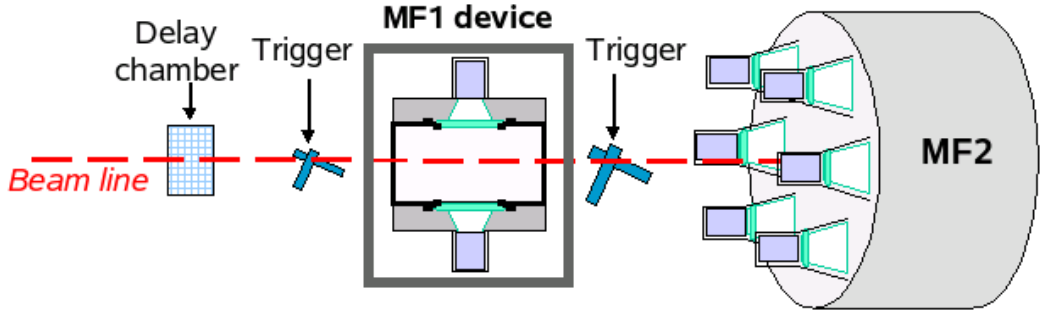}
  \caption{Schematic view of the experimental devices along the test-beam line.}
  \label{fig-Beam-line}
\end{figure}

The SPS-beam line is a pulsed beam delivering about 10\,000
particles by spill (every 16.8\,s). The purity of the beam is better
than 98\% for 50\,GeV electrons.
Along the beam line, upstream MF2, we have installed also the MF1 device,
a beam-position chamber (delay chamber) with a resolution of about 0.6\,mm
and a trigger system composed of two sets of scintillating counters,
one upstream and one downstream the MF1 device, as shown in
Figure~\ref{fig-Beam-line}.
The beam-spot size measured by the delay
chamber is about $4\times7$\,mm$^2$.
%The MF2 chamber, located after the last trigger counter, is aligned with the
%beam line, in order to receive electrons in the center of the pre-shower system.

An external gas system allows to fill both chambers, MF1 and MF2,
with the desired gas mixture, at a specific pressure. The two chambers are
equipped with pressure and temperature gauges to control both parameters during
the data taking. In this paper, we report the FLY measurement of
the following gas mixture: 80\%(N$_{2}$) and 20\%(O$_{2}$).
This composition is close to the atmospheric dry air which is an admixture:
78.08\%(N$_2$)-0.93\%(Ar)-20.99\%(O$_2$). The gas system enables
to fill both chambers at the same time with the same gas,
leading to simultaneous identical measurements with MF1 and MF2.

\section{Pre-shower system}
When interacting in materials, high-energy electrons develop
electromagnetic showers. The longitudinal development and the
lateral spread of the shower are characterized respectively
by the radiation length, $X_{0}$, and by the Moli\`ere radius,
$R_{Mo}$, of the material~\cite{pdg}.
In the air, an electromagnetic shower takes several kilometers for
developing (at atmospheric pressure $X_{0}\simeq 300$\,m). Then, in
order to sample air-shower development in laboratory, we
have to use a fast and compact shower initiator. Here we use a
pre-shower system made of a variable thickness copper target.

We choose copper because of its properties: high density
($\rho$\,=\,8.96\,g/cm$^3$) and low atomic number (Z\,=\,29) which
initiate compact showers with a small lateral spread
($X_{0}$\,=\,14.3\,mm, $R_{Mo}$\,=\,14.9\,mm). The
characteristics of the electromagnetic shower induced in the copper
target are close to the EAS characteristics. The
critical energy is of the same order of magnitude ($\sim$\,24\,MeV in
copper and $\sim$\,80\,MeV in air) and the energy of the secondary
particles is similar. Moreover, the particle density of the shower
in the MF2-chamber gas after the pre-shower (from $10^{3}$ to
$10^{5}$\,e$^\pm$/m$^{2}$) is in the range of the particle density encountered
in UHECR EAS at the shower maximum (from $10^{3}$ to
$10^{7}$\,e$^\pm$/m$^{2}$)~\cite{Risse}.

\begin{figure}
\begin{center}
  \includegraphics[width=12cm]{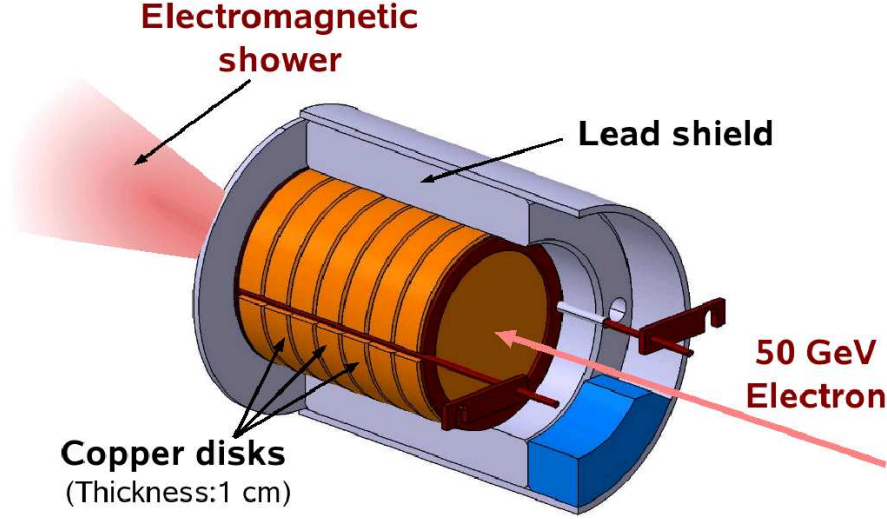}
  \caption{The MF2 pre-shower system is a copper-disk-stack target
  surrounded by a lead-shielding tube.}
  \label{fig-Preshower}
\end{center}
\end{figure}

Figure~\ref{fig-Preshower} shows a sketch of the pre-shower system.
The target is made of a stack of copper disks, 10\,mm thick each. The
age of the shower changes as a function of the number N of disks in
the stack. The equation~\ref{eq-Xn} gives the pre-shower thickness
(expressed in $X_{0}$ units) as a function of N. One copper disk
corresponds to $(0.7\pm 0.002)\,X_{0}$. All the matter on the beam
line upstream the copper target (Trigger scintillator, MF1 chamber,
etc.) corresponds to $(0.27\pm 0.05)\,X_{0}$.
\begin{equation}\label{eq-Xn}
X_{N}=(0.27 + N\times 0.7) X_{0}~.
\end{equation}
In copper, at 50\,GeV, the maximum of the shower development is at 7\,$X_{0}$
($\sim$10\,cm). The pre-shower system allows to reproduce the air shower
development in real atmosphere, on several kilometers, until the shower maximum.
The sampling values used here are: 0, 1, 3, 5, 7 and 10 copper disks.

In order to minimize the background induced in the PMTs by the showers,
the copper-disk-stack target is surrounded by a lead
shielding (20\,mm thick) which protects the PMTs from the
backscattered particles.

\section{Data taking and FLY reconstruction}
%The recording of the events is done on an event per event basis.
The data recording is performed on an event by event basis.
It uses a VME based DAQ system running a Labview program. The signal
from all PMTs (MF1, MF2 and triggers) is recorded by QADC
(CAEN-V792) which integrate the charge during 100\,ns.
We define two kinds of event: Beam Events (BE) and Random Event
(RE). A BE is triggered when an electron passes through the two
sets of scintillator counters (FLY measurement). The RE are randomly
triggered (background measurement). For every run about one million
events are recorded: 500\,000 BE and 500\,000 RE.

The air FLY is rather weak and the majority of photons
emitted are lost in the chamber. The typical mean number of
photon detected by a PMT is about 0.01\,pe/evt (photoelectron per event).
The method to extract the mean Detected Light (DL) of a run from the
data is described in the MF1 paper~\cite{MF1}. This method can reconstruct
the DL at the level as low as 0.01\,pe/evt with an
uncertainty smaller than~4\%.

The detected light (DL) could come from several sources:
Fluorescence (FDL), Cherenkov (CDL) and Background (Bgd).
The overall signal is then:
\begin{equation}
    DL=FDL+CDL+Bgd~.
\end{equation}

\begin{figure}[htb]
 \includegraphics[width=\columnwidth]{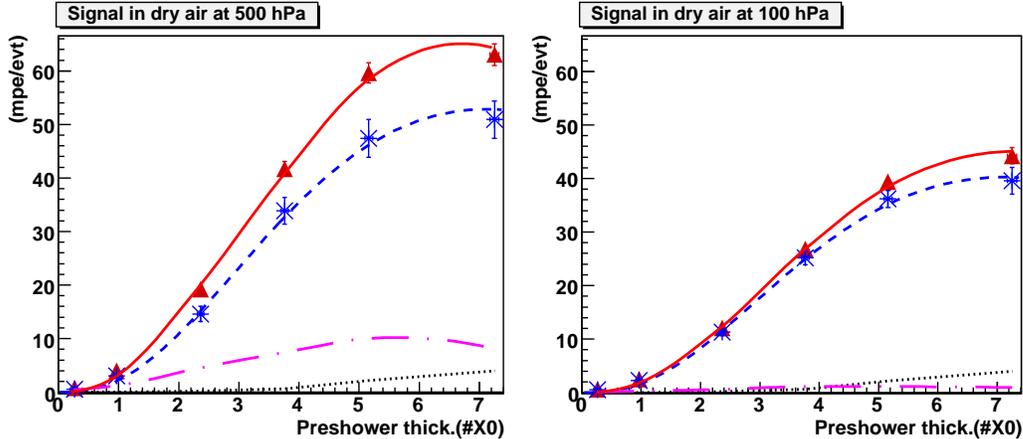}
 \caption{Measured light in dry air at 500\,hPa (left) and 100\,hPa
 (right) in milli-photoelectron per event (mpe/evt) as a function of
 the pre-shower thickness (in $X_0$). Triangles represent the total signal (DL);
 dotted line is for the Bgd estimation from vacuum measurements;
 dot-dashed curve is the CDL simulation; stars are the FDL data
 (after substraction of Bgd and CDL); dashed line is the FLY model
 for showers. The solid line is the sum of the all contributions.}
  \label{fig-DL-composition}
\end{figure}

Figure~\ref{fig-DL-composition} shows the DL reconstructed from
data and the estimation of CDL and Bgd contributions to the total
measured light. The FDL is determined by subtracting CDL and Bgd to
DL. The main part of the DL comes from the fluorescence whatever the conditions
(50\,GeV showers in dry air at 500\,hPa for the left panel or at 100\,hPa
for the right one).

The background level is determined from both, RE of the run and BE in vacuum
where no light from neither fluorescence nor Cherenkov is expected.
The background comes mainly from backscattered particles of the showers.
The MF2 device has been designed to minimize it. As one can see
in figure~\ref{fig-DL-composition}, the background is quite low
compared to the fluorescence signal but it grows with the shower age
(pre-shower thickness). That is why we limit our FLY measurements to
the shower maximum. The uncertainty of the background measurement is
about 20$\%$.

We estimate the Cherenkov radiation contribution with a
Geant4~\cite{Geant4} based Monte-Carlo simulation program. The
Cherenkov light yield is important at atmospheric pressure
($\sim$\,20\,ph/m/electron). However it is not contributing so much to the
detected signal because it is mainly emitted in the forward
direction, downstream, where it is absorbed on the black surface of
the chamber. The uncertainty of the CDL after diffusion on the black surface
is estimated at 50$\%$.

Finally we extract the FLY in MF2 from the FDL, after dividing it by
the MF2 efficiency $\varepsilon_{MF2}$ (see next section):

\begin{equation}\label{eq-FLY}
FLY=\frac{DL - CDL - Bgd }{\varepsilon_{MF2}}.
\end{equation}

\section{Calibration and systematic errors}
The Monte Carlo simulation plays a crucial role in the calibration
and data analysis of the MACFLY experiment. The exact geometry and matter
of all parts of MF2 (pre-shower, gas) and of other objects (MF1, scintillator)
which are in the beam line have been reproduced carefully in a Geant4
based simulation program. For the external parts, only the details bigger than
few centimeters are described. The optical properties of the optical system
and of the inner chamber surfaces are also implemented.
The fluorescence emission is assumed isotropic and its geometrical distribution
is assumed to match the deposited energy distribution. The simulation program tracks
all the optical photons until the PMT-photo-detection surface or until their
absorption.

All the phototubes used for the MACFLY experiment are tested and
cross calibrated in laboratory with a test bench using stabilized UV
LED (370\,nm)~\cite{Colin}. Then, to calibrate the MF2 device we use
the MF1 chamber which is well calibrated~\cite{MF1}.
In order to do that, we performed measurement in both chambers
filled with the same gas at same pressure and temperature, and
without pre-shower target in front of MF2 (0~disk). In this
configuration, both chambers measure FLY induced by the same 50\,GeV-electron
tracks. As the two measurements are in the same condition,
the $FLY/E_d$ should be the same in the two chambers.
\begin{equation}
FLY/E_d = \frac{FDL_{MF1}}{\varepsilon_{MF1}  \times E_{d_{MF1}}}
= \frac{FDL_{MF2}}{\varepsilon_{MF2} \times E_{d_{MF2}}}.
\end{equation}
where $FDL_{MF1}$ and $FDL_{MF2}$ are the
fluorescence light measured (in pe/evt) with MF1 and MF2,
$\varepsilon_{MF1}$ and $\varepsilon_{MF2}$ are the light collection
efficiencies of MF1 and MF2,
$E_{d_{MF1}}$ and $E_{d_{MF2}}$ are the energy deposited in the
air inside the measurement chambers of MF1 and MF2.
The collection efficiency of MF1 is much better than the MF2 one.
For the same light yield, the $FDL_{MF1}$ is about 12 times higher than $FDL_{MF2}$.
The ratio between the $E_{d_{MF1}}$ and $E_{d_{MF2}}$
is estimated by Monte Carlo simulation at 0.15.
Then, we determine the MF2 efficiency from the $\varepsilon_{MF1}$:
\begin{equation}
\varepsilon_{MF2} = \frac{FDL_{MF2}}{FDL_{MF1}}\cdot \frac{E_{d_{MF1}}}{E_{d_{MF2}}} \cdot \varepsilon_{MF1} = (7.0 \pm 1.5) 10^{-5} pe/photon.
\end{equation}

The single-electron-track (no pre-shower) air fluorescence produces
a very weak signal in the MF2 PMTs ($<$\,0.001\,pe/evt).
The uncertainty on this measurement is large and induces an absolute calibration
uncertainty of MF2 larger than for MF1 (see detail in table~\ref{table-systematics}).

\begin{table}
\begin{center}
\begin{tabular}{|c|c|c|}
  \hline
  Errors sources & Absolute & relative \\
  \hline
  MF1 calibration& 13.7\% & -\\
  MF1/MF2  & 18\% & -\\
Geometrical distribution & $\sim 3\%$ & $\sim 3\%$ \\
DL reconstruction & $\sim 3\%$ & $\sim 3\%$ \\
CDL Simulation  & $\sim 1.5\%$ & $\sim 1.5\%$ \\
Bgd Measurement & $\sim 1.5\%$ & $\sim 1.5\%$ \\ \hline
\textbf{TOTAL} & \textbf{23.1\%} & \textbf{$\sim$ 4.7\%} \\
  \hline
\end{tabular}
\end{center}
\caption{Systematic uncertainties of MF2 measurements in dry air at
100\,hPa and for 5\,$X_{0}$-thick-pre-shower target.} \label{table-systematics}
\end{table}

The systematic error of $E_{d_{MF1}}$ and $E_{d_{MF2}}$, obtained by Monte Carlo
simulation, is dominated by the air-density uncertainty
(about 1\% at 500\,hPa and 2\% at 100\,hPa). For the inter-calibration measurement,
both chambers are filled with the same gas, then this systematic is negligible.
However, the spacial geometry distribution of the energy deposited in the MF2
chamber changes as a function of the shower age. As the fluorescence emission
should match the deposited energy distribution, the geometrical acceptance
of MF2 could change as a function of the shower age.
The systematic error induced by this effect has been estimated
by simulation to be less than 3\%.

The background and the Cherenkov radiation represent a small fraction
of the raw measured light (see figure~\ref{fig-DL-composition}) and induce
small systematic errors.
The Cherenkov contribution grows with pressure (density) and induces more
uncertainty at higher pressure.
Table~\ref{table-systematics} shows the contribution of the different systematic
effects to the global uncertainty at 100\,hPa and for 5\,$X_{0}$-thick pre-shower.
The systematic errors of relative measurements are rather good ($<$5\%) and grow
only up to $\sim$7.5\% at 500\,hPa.

\section{Result and discussion}

\begin{figure}
\begin{center}
  \includegraphics[width=10 cm]{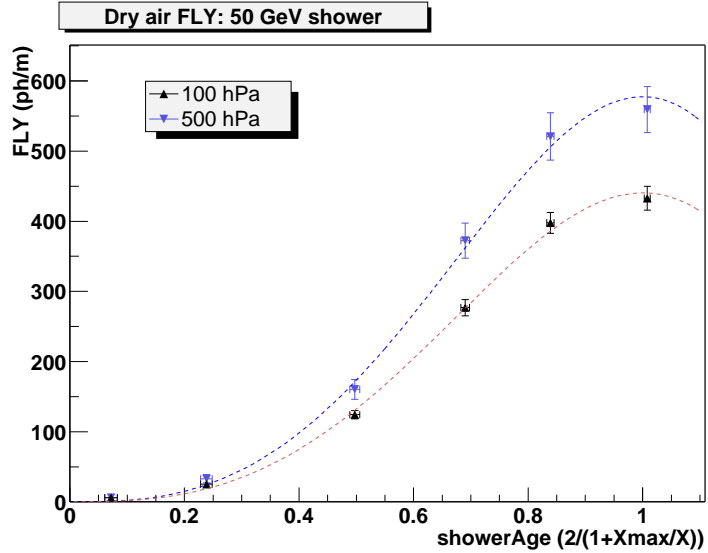}
\caption{Dry-air-Fluorescence-Light Yield per unit of length (photon/meter)
in dry air (100\,hPa \& 500\,hPa) emitted by 50\,GeV electromagnetic
showers as a function of the shower age.
Dotted lines correspond to a model of shower development in copper.}
  \label{fig-FLY-Age-dep}
\end{center}
\end{figure}

We measured FLY of dry air excited by electromagnetic showers
for several pressures and for several shower-age values.
Figure~\ref{fig-FLY-Age-dep} shows the mean number of
fluorescence photons emitted when a 50\,GeV shower traverses
a one-meter-thick layer of air as a function of the shower age.
We have performed such measurements for two
different pressures: 100\,hPa and 500\,hPa.
The dotted lines are proportional to a model of energy lost ($dE/dX$)
by a 50\,GeV-electron-induced shower
developing in copper, based on Geant4 simulations.
The measured air FLY follows well the expected shower development.

\begin{figure}
  \includegraphics[width=\columnwidth]{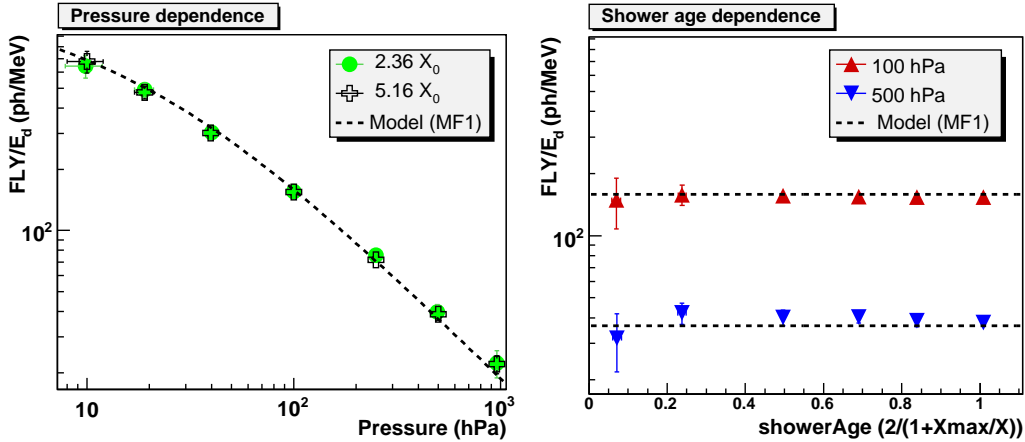}
\caption{Dry-air-Fluorescence-Light Yield per unit of deposited energy
(photon per MeV) as a function of: (left) gas pressure, for two
thicknesses of the pre-shower; (right) shower age, for two values of
the pressure . A comparison with our FLY model~\cite{MF1} is also shown
(dotted lines).}
  \label{fig-FLYonEdep}
\end{figure}

To check the properties of air-fluorescence light induced by air shower,
we compare our results to the air-fluorescence model developed by Colin~\cite{Colin}
found to reproduce well the MF1 results~\cite{MF1}.
This model assumes the air FLY to be proportional to the energy
deposited ($E_d$) in the air volume. Thus, the ratio $FLY/E_d$
(expressed in photons per MeV) should be independent of the
excitation source.
For each FLY measurements with MF2, $E_d$ inside the chamber was obtained by
Monte Carlo simulation. Figure~\ref{fig-FLYonEdep}
shows the variations of $FLY/E_d$ as a function of the pressure
(left panel) and as a function of the shower age (right panel).

The pressure dependence was measured for two pre-shower thickness:
2.36\,$X_{0}$ and 5.16\,$X_{0}$. We compare these results with
the FLY model based on MF1 measurements~\cite{MF1} realized with
the same gas mixture and the same optical filter
(dotted line in Figure~\ref{fig-FLYonEdep}).
For the two shower ages, the air $FLY/E_d$ has the same variation
with pressure as for air fluorescence induced by a single-electron track.
Comparison with other experiments is difficult because it requires a knowledge of
the deposited energy in these experiments and corrections from the experimental differences
(air composition, filter, etc.). However, the MACFLY model has been already compared
with past~\cite{Colin} and recent~\cite{summary} experiments and shows similar pressure dependence.

The shower-age dependence was measured at two pressures: 100\,hPa
and 500\,hPa. In both case, we do not find any significant
variation of $FLY/E_d$ with the shower age, in agreement with the
FLY model (doted lines). This result is also in good agreement with
the FLASH thick-target experiments which measured shower-age dependence
in ambient air~\cite{flash2}.
This shows clearly that the $FLY/E_d$ properties are
independent of the excitation source of the air. There is no clue
of special behavior from saturation effect at the EAS density, or
from low-energy-electron excitation of air molecules as it could
be expected~\cite{Arqueros}. Thus, air-FLY results obtained from
electron-track experiments (thin target) can be directly used to
determine the properties of the fluorescence induced by air showers.
According to our result, the systematic error induced by
the extrapolation from electron track to electromagnetic shower
must be less than 5\%.

\section{Conclusion}

Using the MF2 device of the MACFLY experiment, we measured
the Fluorescence-Light Yield induced by 50\,GeV electromagnetic
showers in a laboratory-controlled air (80\%(N$_{2}$)-20\%(O$_{2}$)).
We studied both pressure and shower-age dependencies. The FLY variations with
pressure, measured at two shower ages, are the same as the one measured
with the single-electron-track device MF1~\cite{MF1}. The FLY variations with
shower age, measured at two pressures, are well reproduced by
the shower-development simulations implementing our air-fluorescence
model which assumes the air FLY to be proportional to the energy deposited
in the air volume. No evidence for FLY variation with the air-excitation
source was found.

\section{Acknowledgments}
The fund for this work has been partly provided by the Institut
National de Physique Nucl\'eaire et de Physique des Particules and
by the Joint Institute for Nuclear Research. We would like to
thank the Centre Europ\'een de Recherche Nucl\'eaire who allocated
two weeks of beam test on the SPS beam line and particularly  the
CERN PH-DT2 (Detector Technology) group. Many thanks to our
colleagues from the Alice experiment who helped and supported our
activity in the test beam area and to M. Maire for his Geant4
expertise.

%\newpage
%\listoffigures

%\newpage
%\listoftables

\end{document}